\documentstyle[amssymb,12pt]{article}
\begin{document}

\title{Deformations of the fermion realization \\ of the sp(4) 
algebra and its subalgebras }
%EndAName
\author{
K. D. Sviratcheva$^{1}$,
A. I. Georgieva$^{1,2}$,
V. G. Gueorguiev$^{1}$, \\
J. P. Draayer$^{1}$ and M. I. Ivanov$^{2}$ \\
\\
$^{1}${\it Department of Physics and Astronomy,} \\
{\it \ Louisiana State University,} \\
{\it \ Baton Rouge, Louisiana 70803 USA} \\
\\
$^{2}${\it Institute of Nuclear Research and Nuclear Energy,}\\
{\it \ Bulgarian Academy of Sciences, Sofia 1784, Bulgaria} \\
}

\date{\today}

\maketitle

\begin{abstract}
With a view towards future applications in nuclear physics, the fermion
realization of the compact symplectic sp(4) algebra and its q-deformed
versions are investigated. Three important reduction chains of the sp(4)
algebra are explored in both the classical and deformed cases. The deformed
realizations are based on distinct deformations of the fermion creation and
annihilation operators. For the primary reduction, the su(2) sub-structure
can be interpreted as either the spin, isospin or angular momentum algebra,
whereas for the other two reductions su(2) can be associated with pairing
between fermions of the same type or pairing between two distinct fermion
types. Each reduction provides for a complete classification of the basis
states. The deformed induced u(2) representations are reducible in the action
spaces of sp(4) and are decomposed into irreducible representations.
\end{abstract}

\section{Introduction}

Symplectic algebras can be used to describe many-particle systems. The compact,
$sp(2n),$ and noncompact versions, $sp(2n,R),$ of the algebra enter naturally
when the number of particles or couplings between the particles change in a
pairwise fashion from one configuration to the next. In this paper we consider
the simplest nontrivial case: the compact $sp(4)$ symplectic algebra which is
isomorphic to the Lie algebra of the five-dimensional rotation group $SO(5)$
\cite{GoLiSp,hecht,bar}. Applications of $sp(4)$ are related to different
interpretations of the quantum numbers of the fermions used to construct the
generators of the $Sp(4)$ group.

Interest in symplectic groups is related to applications to nuclear structure
\cite{Goli,Roro, pehe}. In particular, $Sp(4)$ has been used to explore
pairing correlations in nuclei \cite{GoLiSp,ker,cirevo}. The reduction
chains to different realizations of the $u(2)$ subalgebra of $sp(4)$ yield a
complete classification scheme for the basis states. It is rather easy to
generalize this work to higher rank algebras and therefore the algebraic
techniques are illustrated by the $sp(4)$ example \cite{klmap}. A further
interest in the symplectic algebras is related to their use in mapping
methods from the fermion space to the space spanned by collective bosons and
ideal fermions
\cite{nagedo}. In these applications the primary purpose is to simplify the
Hamiltonian of the initial problem.

In the last decade a lot of effort, from a purely mathematical \cite
{Drinfeld,liNoRu,aach,hay} as well as from the physical point of view
\cite{sha}, has been concentrated on various deformations of the classical
Lie algebras. The general feature of these deformations is that at the limit
of the deformation parameter $q\rightarrow 1,$ the $q$-algebra reverts back
to the classical Lie algebra. More than one deformation can be realized for
one and the same ``classical'' algebra, which can be chosen in a convenient
way in different physical applications. There are many similarities between
the classical Lie algebras and their deformations, especially with respect to
their representation. Deformed algebras introduce a new degree of freedom
that can give a better explanation of non-linear effects. Their study can
lead to deeper understanding of the physical significance of the deformation.

In \cite{debre} a boson realization of the noncompact $sp(4,R)$ and two
distinct deformations of it, as well as compact and noncompact subalgebras of
each, were investigated and reductions of their action spaces obtained. As the
fermion case has more direct application in nuclear theory than the boson
construction, in this work our aim is to investigate in detail the fermion
realization of the $sp(4)$ algebra and its deformations. Using the methodology
from \cite{debre}, we begin with the well-known realization of this algebra in
terms of ``classical'' fermion creation and annihilation operators 
and consider all the subalgebras which correspond to different ways of
specifying labels of the basis states via the eigenvalues of the operators
generating these subalgebras (Section 2). Furthermore, we obtain the
deformation of this
$sp(4)$ algebra and its subalgebras by introducing a transformation function
that deforms the classical fermions into $q$-deformed fermions. We
also introduce another deformation in terms of the standard $q$-fermions
and by following the same procedure we investigate the enveloping algebra
of
$sp(4)$ and the action of its generators on the deformed basis (Section 3).

\section{Fermion realization of the sp(4) algebra}

To establish the notation, recall some features of the fermion realization of
the $sp(4)$ algebra \cite{bar,Goli, 
klmap}, which is isomorphic to $so(5)$ \cite{hecht}, as normally used in the
shell-model studies. The operator
$c_{m,\sigma }^{\dagger }$ creates ($c_{m,\sigma }$ annihilates) a particle of
type $\sigma =\pm 1,$ in a state of total angular momentum $j=\frac{2k+1}{2},\
k=0,1,2...,$  with projection $m$ along the $z$ axis ($-j\leq m\leq
j$ ). These operators satisfy Fermi anticommutation relations:

\begin{equation}
\begin{array}{ll}
\{c_{m^{\prime },\sigma ^{\prime }},c_{m,\sigma }^{\dagger }\}=\delta
_{m^{\prime },m}\delta _{\sigma ^{\prime },\sigma }, & \{c_{m^{\prime
},\sigma ^{\prime }}^{\dagger },c_{m,\sigma }^{\dagger }\}=\{c_{m^{\prime
},\sigma ^{\prime }},c_{m,\sigma }\}=0,
\end{array}
\label{cfc}
\end{equation}
and Hermitian conjugation is given by $(c_{m,\sigma }^{\dagger
})^{*}=c_{m,\sigma }.$

For a given $\sigma ,$ the dimension of the fermion space is $2\Omega
_{j}=2j+1.$ The fermion realization of $sp(4)$ is given in a standard way by
means of the following operators \cite{GoLiSp,hecht, klmap}:

\begin{eqnarray}
{{{A_{\sigma ,\sigma ^{\prime }}}}} &=&\xi _{\sigma ,\sigma ^{\prime
}}\sum_{m=-j}^{j}{{(-1)}^{j-m}}c_{m,\sigma }^{\dagger }c_{-m,\sigma ^{\prime
}}^{\dagger }{=}A{{{_{\sigma ^{\prime },\sigma }=(B_{\sigma ,\sigma ^{\prime
}})}}}^{\dagger },  \label{prlg} \\
{{{B_{\sigma ,\sigma ^{\prime }}}}} &=&\xi _{\sigma ,\sigma ^{\prime
}}\sum_{m=-j}^{j}{{(-1)}^{j-m}}c{_{-m,\sigma }}c{_{m,\sigma ^{\prime }}={{\
B_{\sigma ^{\prime },\sigma }=(A_{\sigma ,\sigma ^{\prime }})}}}^{\dagger }.
\label{palg}
\end{eqnarray}
These operators create (annihilate) a pair of fermions coupled to total
angular momentum $J=0$ \cite{hecht} and thus constitute
boson-like objects according to the Spin-Statistics theorem
\cite{spin-staistics theorem} when the operators

\begin{equation}
D_{\sigma ,\sigma ^{\prime }}=\eta \sum_{m=-j}^{j}c{{{_{m,\sigma }^{\dagger }
}}}c{_{m,\sigma ^{\prime }}},  \label{cg}
\end{equation}
preserve the number of fermions. Here the normalization constants are

\begin{equation}
\xi _{\sigma ,\sigma ^{\prime }}=\frac{\eta }{\sqrt{(1+\delta _{\sigma
,\sigma ^{\prime }})}},\eta =\frac{1}{\sqrt{2\Omega _{j}}}.
\label{shell scalling constants}
\end{equation}

The number of the operators $A_{\sigma ,\sigma ^{\prime }},B_{\sigma
,\sigma ^{\prime }}$ and ${D_{\sigma ,\sigma ^{\prime }}}$ is ten $(
A_{\sigma ,\sigma ^{\prime }}=A_{\sigma ^{\prime },\sigma },$ $\
B_{\sigma ,\sigma ^{\prime }}={{B_{\sigma ^{\prime },\sigma }}})$. Their
commutation relations, obtained by means of (\ref{cfc}), show that these
operators generate a fermion realization of the $sp(4)$ algebra
\cite{GoLiSp}. An additional index $m\neq 0$ of the creation and annihilation
fermion operators is introduced in order to construct non-zero operators
${{{A} _{\sigma ,\sigma }}}$ and ${{{B_{\sigma ,\sigma }}}}${,} but the index
$
\sigma =\pm 1$ defines the algebraic properties of the generators ${{{
A_{\sigma ,\sigma ^{\prime }},B_{\sigma ,\sigma ^{\prime }}}}}$ and ${
D_{\sigma ,\sigma ^{\prime }}.}$

Different interpretations of $\sigma $ correspond to different
physical meanings for the operators generating the ten-parametric $Sp(4)$
group and therefore different physical models. These can be used to describe
various aspects of the nuclear interaction (different Hamiltonians)\cite{klmap}
like charge independent pairing, two level pairing (Lipkin model) or two
dimensional rotations and vibrations. The $sp(4)$ algebra is considered to be
the dynamical symmetry algebra in these applications. Each of the limits is
described by a reduction chain of the algebra which serves to label the basis
states by eigenvalues of the invariant operators of the subalgebras and gives
the corresponding limiting forms of the model Hamiltonian. 

\subsection{Subalgebras of sp(4)}

The investigation of the subalgebras of $sp(4)$ contained in its reduction
chains is given below.

\begin{enumerate}
\item  By using the particle number preserving Weyl generators $D_{i,j}$ (%
\ref{cg}), a subalgebra $u(2)$ of $sp(4)$ is realized by the operators:
\begin{equation}
\begin{array}{ll}
{{\tau }_{1}\equiv D_{+1,-1},} & {{\tau }_{0}}=\frac{N{_{1}}-N{_{-1}}}{2},
\\ {{\tau }_{-1}\equiv D_{-1,+1},} & {N}={N_{+1}}+{N_{-1},}
\end{array}
\label{psg}
\end{equation}
where ${N_{\pm 1}\equiv }\frac{1}{\eta }{D_{\pm 1,\pm 1}}$ are the operators
of the total number of fermions of each kind,
\begin{equation}
N_{\sigma }{=}\sum_{m=-j}^{j}c{{{_{m,\sigma }^{\dagger }}}}c{_{m,\sigma }\ ,\
}
\sigma =\pm 1.  \label{Nc}
\end{equation}
The action of these operator on the fermion creation and annihilation
operators is given by

\begin{equation}
\begin{array}{ll}
N_{\sigma }c_{m,\sigma ^{\prime }}^{\dagger }=c_{m,\sigma ^{\prime
}}^{\dagger }(N_{\sigma }+\delta _{\sigma ,\sigma ^{\prime }})\ , & N_{\sigma
}c_{m,\sigma ^{\prime }}=c_{m,\sigma ^{\prime }}(N_{\sigma }-\delta _{\sigma
,\sigma ^{\prime }})
\end{array}
,  \label{NcCom}
\end{equation}
\[
\ \sigma ,\sigma ^{\prime }=\pm 1,
\]

and the anticommutation relations (\ref{cfc}) yield the equality
\begin{equation}
\sum_{m=-j}^{j}c_{m,\sigma }c_{m,\sigma }^{\dagger }=2\Omega _{j}-N_{\sigma
}\ ,\ \sigma =\pm 1.  \label{Nc-}
\end{equation}

The operators (\ref{psg}) satisfy the $u(2)$ commutation relations
\begin{equation}
\begin{tabular}{llll}
${\lbrack }\tau _{+},\tau _{-}{]}=2\frac{\tau _{0}}{2\Omega _{j}},$ & ${[}%
\tau _{0},\tau _{\pm }{]}=\pm \tau _{\pm },$ & $\left[ N,\tau _{\pm }\right]
=0,$ & $\left[ N,\tau _{0}\right] =0,$
\end{tabular}
\label{crc}
\end{equation}

where $\tau _{0},\tau _{\pm }$ close on an algebra $su^{\tau }(2)\ $
that is isomorphic to $so(3).$ The operator $N$ generates $U(1)$ and plays
the role of the first order invariant of $U^{\tau }(2)=SU^{\tau }(2)\otimes
U(1)$. The second order Casimir operator of $SU^{\tau }(2)$ is
given by
\begin{equation}
{\bf \tau }^{2}=\frac{2\Omega _{j}}{2}(\tau _{+}\tau _{-}+\tau _{-}\tau
_{+})+\tau _{0}\tau _{0},  \label{i2}
\end{equation}

and the second order invariant of $U^{\tau }(2)$ \cite{klmap} is simply
\begin{equation}
C_{2}=N(N+1)-{\bf \tau }^{2}.
\end{equation}

The algebra $su^{\tau }(2)\backsim so(3)$ plays a very important role in all
kinds of different physical applications since it is of the standard spin
type, which can be interpreted as spin, isospin or angular momentum in the
various models.

\item  Another unitary realization of $u(2),$ denoted by $u^{0}(2),$ is
generated by $\tau _{0}$ (\ref{psg}) and the operators
\begin{equation}
A_{+1}^{0}\equiv {{{A_{1,-1},\quad }}}A_{-1}^{0}\equiv {{{B_{1,-1},\quad }}%
}A_{0}^{0}\equiv \frac{N}{2}-\Omega _{j}  \label{gu0}
\end{equation}
with the following commutation relations:

\begin{eqnarray}
\lbrack A_{+1}^{0},A_{-1}^{0}] &=&2\frac{A_{0}^{0}}{2\Omega _{j}},\qquad
\left[ A_{0}^{0},A_{\pm 1}^{0}\right] =\pm A_{\pm 1}^{0}  \label{crAzero} \\
\lbrack \tau _{0},A_{\pm 1}^{0}] &=&0,\qquad [\tau _{0},A_{0}^{0}]=0.
\nonumber
\end{eqnarray}

For this realization the operator $\tau _{0}$ acts as a first order
invariant of $u^{0}(2)$, defining the reduction $u^{0}(2)=su^{0}(2)\oplus
u^{0}(1).$ The second order Casimir invariant of this subgroup is given as
\begin{equation}
C_{2}(SU^{0}(2))=\frac{2\Omega _{j}}{2}%
(A_{+1}^{0}A_{-1}^{0}+A_{-1}^{0}A_{+1}^{0})+A_{0}^{0}A_{0}^{0}.  \label{CSU0}
\end{equation}
The generators of this $SU^{0}(2)$ group are operators pairing particles of
two different kinds.

\item  Next, we consider two mutually complementary $su\left( 2\right) $
subalgebras of the algebra $sp(4),$ denoted by $su^{+}(2)$ and $su^{-}(2)$.
These algebras are generated by the operators
\begin{equation}
A_{+1}^{\pm }\equiv A_{\pm 1,\pm 1},\quad B_{-1}^{\pm }\equiv B_{\pm 1,\pm
1},\quad \ D_{0}^{\pm }\equiv \frac{N_{\pm 1}}{2}-\frac{\Omega _{j}}{2},
\label{cgpm}
\end{equation}
with the following commutation relations:
\begin{equation}
\begin{tabular}{lll}
$\left[ A_{+1}^{\pm },B_{-1}^{\pm }\right] =4\frac{D_{0}^{\pm }}{2\Omega _{j}
},$ & $\left[ D_{0}^{\pm },A_{+1}^{\pm }\right] =A_{+1}^{\pm },$ & $\left[
D_{0}^{\pm },B_{-1}^{\pm }\right] =-B_{-1}^{\pm }.$%
\end{tabular}
\label{cpmc}
\end{equation}
It is simple to see that each of the generators of $SU^{+}(2)$ commutes with
all of the $SU^{-}(2)$ generators. The second order Casimir operators of the
$SU^{\pm }(2)$ are
\begin{equation}
C_{2}(SU^{\pm }(2))=\frac{\Omega _{j}}{2}(A_{+1}^{\pm }B_{-1}^{\pm }+B_{-1}^{%
\pm }A_{+1}^{\pm })+D_{0}^{\pm }D_{0}^{\pm }.  \label{CSU+-}
\end{equation}
In this case the addition of the operators $N_{\mp 1},$ considered to be
generators of the subgroups $U^{\mp }(1)$, extend $su^{\pm }(2)$ to $u^{\pm
}(2)=$ $su^{\pm }(2)$ $\oplus u^{\mp }(1)$. $N_{\mp 1}$ act as first order
Casimir operators of $U^{\pm }(2).$ The operators closing the two mutually
complementary subalgebras describe pairs of particles of the same kind.

\item  The sum of the generators of the groups $SU^{+}(2)$ and $SU^{-}(2)$
give rise to another unitary realization of $su\left( 2\right) $ subalgebra
of $sp\left( 4\right) $ denoted by $su^{\symbol{126}}(2)$,
\begin{equation}
\tilde{A}_{+1}\equiv A_{+1}^{+}+A_{+1}^{-},\quad \tilde{B}_{-1}\equiv
B_{-1}^{+}+B_{-1}^{-},\quad \ \tilde{D}_{0}\equiv \frac{N_{1}}{2}+\frac{
N_{-1}}{2}-\Omega _{j},
\end{equation}
with the following commutation relations:
\begin{equation}
\begin{tabular}{lll}
$\left[ \tilde{A}_{+1},\tilde{B}_{-1}\right] =4\frac{\tilde{D}_{0}}{2\Omega
_{j}},$ & $\left[ \tilde{D}_{0},\tilde{A}_{+1}\right] =\tilde{A}_{+1},$ & $
\left[ \tilde{D}_{0},\tilde{B}_{-1}\right] =-\tilde{B}_{-1},$%
\end{tabular}
\end{equation}
and the second order Casimir invariant
\begin{equation}
{C_{2}}(SU^{\symbol{126}}(2))=\frac{\Omega _{j}}{2}(\tilde{A}_{+1}\tilde{B}
_{-1}+\tilde{B}_{-1}\tilde{A}_{+1})+\tilde{D}_{0}\tilde{D}_{0}.
\end{equation}
\end{enumerate}

\subsection{Action space of the fermion realization of sp(4)}

In general, the classical fermion operators act in a finite space{\rm %
\ }${\cal E}_{j}$ for a particular $j$-level. The finite representation is
due to the Pauli principle, $c{{{_{m,\sigma }^{\dagger }}}}c{{{_{m,\sigma
}^{\dagger }}}}|0\rangle =0,$ that allows no more than $2\Omega _{j}$
identical fermions in a single $j$-shell. In ${\cal E}_{j}$ the vacuum
$|0\rangle $ is defined by $c_{m,\sigma }$ $|0\rangle =0$ and the scalar
product is chosen so that $\langle 0|0\rangle =1.$

The states that span the ${\cal E}_{j}$ spaces consist of different numbers
of fermion creation operators acting on the vacuum state.  These satisfy the
Pauli principle through their anti-commutation relations (\ref{cfc}). They
form an orthonormal basis in each space and are the common eigenvectors of the
fermion number operators $N_{1}$, $N_{-1}$ $(N_{\sigma }=N_{\sigma }^{*}, $
$\sigma =\pm 1)$ \ and $N=N_{1}+N_{-1}.$ In this way, they span two subspaces
${\cal E}_{j}^{\pm }$ labeled by the eigenvalue of the invariant operator
$P=(-1)^{N}$ of $Sp(4)$. Here we are interested in the even space $ {\cal
E}_{j}^{+},$ containing states of coupled fermions, in order to apply the
theory to the phenomena like pairing correlations in nuclei.

If we introduce
\begin{equation}
{A}_{\frac{1}{2}(\sigma +\sigma ^{\prime })}^{\dagger }{{{\equiv A_{\sigma
,\sigma ^{\prime }},\quad B_{-\frac{1}{2}(\sigma +\sigma ^{\prime })}{{{
\equiv B_{\sigma ,\sigma ^{\prime }},\quad }}}}}}\sigma ,\sigma ^{\prime
}=\pm 1,  \label{prlg2}
\end{equation}
for operators creating (\ref{prlg}) and annihilating (\ref{palg}) a pair of
particles, it is easy to check that they are components of two conjugated
vectors $ \left\{ {A}_{k}^{\dagger }\right\} _{k=0,\pm 1}$ and $\left\{ {B}
_{-k}\right\} _{k=0,\pm 1}$, $k=\frac{1}{2}{(}\sigma +\sigma ^{\prime
})=0,\pm 1$ with respect to the subgroup $SU^{\tau }(2)$
(\ref{psg},\ref{crc}):

\begin{equation}
\begin{array}{ll}
\lbrack \tau _{0},{A}_{k}^{\dagger }]=k{A}_{k}^{\dagger }, & \ [\tau _{l},{A}
_{k}^{\dagger }]=\frac{1}{\sqrt{\Omega _{j}}}{A}_{l+k}^{\dagger },\quad
l=\pm 1, \\
\lbrack \tau _{0},{B}_{k}]=k{B}_{k}, & \ [\tau _{l},{B}_{k}]=-\frac{1}{\sqrt{
\Omega _{j}}}{B}_{l+k},\quad l=\pm 1.
\end{array}
\end{equation}
In the models where ${\bf \tau }$ is interpreted as the isospin operator, ${A
}_{0,\pm 1}^{\dagger }$ $\left\{ {B}_{0,\mp 1}\right\} $ create (destroy) a
pair of fermions coupled to a total isospin $\tau =1$.

Thus, a linearly independent set of vectors that span the ${\cal E}
_{j}^{+}$ space can be expressed in terms of the `boson creation operators'
acting on the vacuum state,
\begin{equation}
\left| \Omega _{j};n_{1},n_{0},n_{-1}\right) =\left( {{{A_{1}^{\dagger }}}}
\right) ^{n_{1}}\left( {{{A_{0}^{\dagger }}}}\right) ^{n_{0}}\left( {{{
A_{-1}^{\dagger }}}}\right) ^{n_{-1}}\left| 0\right\rangle .  \label{csA}
\end{equation}
The basis is obtained by orthonormalization of (\ref{csA}). The operators ${A
}_{0,\pm 1}^{\dagger }$ commute among themselves and therefore form a
symmetric representation. The eigenvalue of the second order Casimir operator
$\Omega _{j}$ labels each representation of $Sp(4)$,
\begin{eqnarray}
C_{2}\left( Sp(4)\right) &=&\left\{ \tau _{+},\tau _{-}\right\} +\left\{
A_{+1}^{0},A_{-1}^{0}\right\} +\left\{ A_{+1}^{+},B_{-1}^{+}\right\}
+\left\{ A_{+1}^{-},B_{-1}^{-}\right\} +  \nonumber \\
&&+\frac{1}{\Omega _{j}}\left( \tau _{0}\tau _{0}+\frac{(N-2\Omega _{j})^{2}%
}{4}\right) ,
\end{eqnarray}
\begin{equation}
C_{2}\left( Sp(4)\right) \left| \Omega _{j};n_{1},n_{0},n_{-1}\right)
=\left( \Omega _{j}+3\right) \left| \Omega _{j};n_{1},n_{0},n_{-1}\right) .
\end{equation}
The group $Sp(4)$ is of rank two and thus there exist two invariant
operators that commute with all the generators of the group \cite{GoLiSp}.
The other invariant operator is of fourth order and it is linearly dependent
on the Casimir operator for this group. Usually representations of $Sp(4)$ are
labeled by the largest eigenvalue of the number operator $N$ and the reduced
isospin of the uncoupled fermions in the corresponding state
\cite{hecht,flo}. In each representation of $Sp(4)$ in the vector space
spanned over (\ref{csA}), the maximum number of particles is $4\Omega _{j}\ $
and the respective state consists of no uncoupled fermions (reduced isospin
zero). It follows that only one quantum number is needed, $\Omega _{j}.$
Within a representation, $\Omega _{j}$ is dropped from the labelling of the
states. Another consequence of the symmetric representation is that the
vector space consists of states of a system with total angular momentum $
J=0^{+}$.

Each representation labelled by $\Omega _{j}$ is finite, because of the
fermion structure of the operators ${A}_{0,\pm 1}^{\dagger }$: $\left( {{{
A_{\pm 1}^{\dagger }}}}\right) ^{\Omega _{j}+1}\left| 0\right\rangle =0$ or $
\left( {{{A_{\pm 1}^{\dagger }}}}\right) ^{\Omega _{j}}\left( {{{
A_{0}^{\dagger }}}}\right) \left| 0\right\rangle =0.$ Another consequence of
the fermion realization is that some of the vectors (\ref{csA}) of the
finite space ${\cal E}_{j}^{+}$are linearly dependent, for example $\left( {{
{A_{1}^{\dagger }}}}\right) ^{\Omega _{j}}\left( {{{A_{-1}^{\dagger }}}}
\right) ^{\Omega _{j}}\left| 0\right\rangle \sim \left( {{{A_{0}^{\dagger }}}
}\right) ^{2\Omega _{j}}\left| 0\right\rangle .$

The states (\ref{csA}) are the common eigenvectors of the fermion number
operators $N_{1}$, $N_{-1}$ $(N_{\sigma }=N_{\sigma }^{*},$ $\sigma =\pm 1):$
\begin{eqnarray}
N_{1}\left| n_{1},n_{0},n_{-1}\right) &=&(2n_{1}+n_{0})\left|
n_{1},n_{0},n_{-1}\right) ,  \nonumber \\
N_{-1}\left| n_{1},n_{0},n_{-1}\right) &=&(2n_{-1}+n_{0})\left|
n_{1},n_{0},n_{-1}\right) ,
\end{eqnarray}
or of the operators $N$ $=N_{1}+N_{-1}$ and $\tau _{0}$\ $=\frac{1}{2}
(N_{1}-N_{-1})$ which are both diagonal in the basis (\ref{csA}):
\begin{eqnarray}
N\left| n_{1},n_{0},n_{-1}\right) &=&n\left| n_{1},n_{0},n_{-1}\right)
,n=2(n_{1}+n_{-1}+n_{0}).  \label{nev} \\
\tau _{0}\left| n_{1},n_{0},n_{-1}\right) &=&i\left|
n_{1},n_{0},n_{-1}\right) ,i=n_{1}-n_{-1}.  \label{iev}
\end{eqnarray}
Their eigenvalues can be used to classify the basis within a representation $%
\Omega _{j}$. The basis states labeled by $\left| n_{1},n_{0},n_{-1}\right) $
for $\Omega _{3/2}=2$ are shown in {\bf Table 1}, where $n$ enumerates
the rows and $i$ the columns.

\[
\begin{tabular}{||c||cc|c|cc}
\hline\cline{2-6}
$n$/$i$ & 2 & \multicolumn{1}{|c|}{1} & 0 & -1 & \multicolumn{1}{|c|}{-2} \\
\hline\cline{2-6}
0 &  &  & $|0,0,0)$ &  &  \\ \cline{1-1}\cline{3-5}
2 &  & \multicolumn{1}{|c|}{$|1,0,0)$} & $|0,1,0)$ & $|0,0,1)$ &
\multicolumn{1}{|c}{} \\ \hline
4 & $|2,0,0)$ & \multicolumn{1}{|c|}{$|1,1,0)$} &
\begin{tabular}{l}
$|1,0,1)$ \\
$|0,2,0)$%
\end{tabular}
& $|0,1,1)$ & \multicolumn{1}{|c|}{$|0,0,2)$} \\ \hline
6 &  & \multicolumn{1}{|c|}{
\begin{tabular}{l}
$|2,0,1)$ \\
$|1,2,0)$%
\end{tabular}
} &
\begin{tabular}{l}
$|1,1,1)$ \\
$|0,3,0)$%
\end{tabular}
& $
\begin{tabular}{l}
$|1,0,2)$ \\
$|0,2,1)$%
\end{tabular}
\ $ & \multicolumn{1}{|c}{} \\ \cline{1-1}\cline{3-5}
8 &  &  &
\begin{tabular}{l}
$|2,0,2)$ \\
$|1,2,1)$ \\
$|0,4,0)$%
\end{tabular}
&  &  \\ \cline{1-1}\cline{4-4}
\end{tabular}
\]

The basis vectors are degenerate in the sense that more than
one of the common eigenstates of the operators $N$ and $\tau _{0}$ have one
and the same eigenvalues $\{n,$ $i\}$ and thus belong to one and the same
cell of {\bf Table 1}. They can be distinguished as eigenstates of the
Casimir operators of the limiting cases:

\begin{enumerate}
\item  The basis states can be labeled by the eigenvalues of the invariant
operator of each subgroup in the reduction ${Sp(4)\supset {U(2)}\supset SU}
^{\tau }{(2)}\otimes {U_{N}(1).}$ As a first order invariant of ${U(2)}$,
the operator $N$ decomposes the spaces ${\cal E}_{j}^{+}$ into a direct sum
of eigensubspaces, defined by the condition that $n$ is fixed (\ref{nev}),
\begin{equation}
n=0,2,4,....,4\Omega _{j}.
\end{equation}

So an irreducible unitary representation ($IUR$) of $U(2)$ is
realized in each row of {\bf Table 1.} The $SU^{\tau }(2)$ subgroup provides
the other two quantum numbers as a standard label of the basis vectors.
First, it is the eigenvalue of the Casimir operator of second rank in ${\bf
\tau },$
\begin{equation}
{\bf \tau }^{2}|n,\tau ,i\rangle ={\tau }({{\tau }+1)}|n,\tau ,i\rangle ,
\end{equation}
where ${\tau =}\frac{\widetilde{n}}{2},$ $\frac{\widetilde{n}}{2}-2,...,1$
(odd) or $0$ (even)$,$ $\widetilde{n}=\min \{n,4\Omega _{j}-n\}$, and second
it is the eigenvalue of $\tau _{0}$ (\ref{iev}), where $i=n_{1}-n_{-1}=-\tau
,-\tau +1,...,\tau.$ As an example, the orthonormalized basis
$ |n,\tau ,i\rangle $ given in terms of the states $\left|
n_{1},n_{0},n_{-1}\right) $ is shown in {\bf Table 2} for $\Omega _{3/2}=2$
{\bf .}
\[
\begin{tabular}{|c|cc|c|cc}
\hline\hline
{\tiny $n_{(\tau )}/i$} & {\tiny 2} & \multicolumn{1}{|c|}{\tiny 1} & {\tiny %
0} & {\tiny -1} & \multicolumn{1}{|c|}{\tiny -2} \\ \hline\hline
$
{\tiny \begin{array}{l}
0 \\
\stackrel{\tau =0}{\rightarrow }
\end{array}}
$ &  &  & {\tiny $|0,0,0)$} &  &  \\ \cline{1-1}\cline{3-5}
$
{\tiny \begin{array}{l}
2 \\
\stackrel{\tau =1}{\rightarrow }
\end{array}}
$ &  & \multicolumn{1}{|c|}{\tiny $|1,0,0)$} & {\tiny $|0,1,0)$} & {\tiny $%
|0,0,1)$} & \multicolumn{1}{|c}{} \\ \hline
$
{\tiny \begin{array}{l}
\stackrel{\tau =2}{\rightarrow } \\
4 \\
\stackrel{\tau =0}{\rightarrow }
\end{array}}
$ &
\begin{tabular}{c}
{\tiny $|2,0,0)$} \\
\\
\\
\end{tabular}
& \multicolumn{1}{|c|}{
\begin{tabular}{c}
{\tiny $\sqrt{2}|1,1,0)$} \\
\\
\\
\end{tabular}
} &
\begin{tabular}{l}
{\tiny $\frac{|0,2,0)}{\sqrt{2}}+\frac{|1,0,1)}{\sqrt{2}}$} \\
\\
{\tiny $\frac{|0,2,0)}{\sqrt{5}}-\frac{2|1,0,1)}{\sqrt{5}}$}
\end{tabular}
&
\begin{tabular}{c}
{\tiny $\sqrt{2}|0,1,1)$} \\
\\
\\
\end{tabular}
& \multicolumn{1}{|c|}{
\begin{tabular}{c}
{\tiny $|0,0,2)$} \\
\\
\\
\end{tabular}
} \\ \hline
$
{\tiny \begin{array}{l}
6 \\
\stackrel{\tau =1}{\rightarrow }
\end{array}}
$ &  & \multicolumn{1}{|c|}{$
{\tiny \begin{array}{c}
|2,0,1)\equiv  \\
-2|1,2,0)
\end{array}}
$} & $
{\tiny \begin{array}{c}
2|1,1,1)\equiv  \\
\frac{-2}{3}|0,3,0)
\end{array}}
$ & $
{\tiny \begin{array}{c}
|1,0,2)\equiv  \\
-2|0,2,1)
\end{array}}
\ $ & \multicolumn{1}{|c}{} \\ \cline{1-1}\cline{3-5}
$
{\tiny \begin{array}{l}
8 \\
\stackrel{\tau =0}{\rightarrow }
\end{array}}
$ &  &  & $
{\tiny \begin{array}{c}
\frac{2}{3}|0,4,0)\equiv  \\
-2|1,2,1)\equiv  \\
|2,0,2)
\end{array}}
$ &  &  \\ \cline{1-1}\cline{4-4}
\end{tabular}
\]

$\ $In general, the state with the maximum number of particles always has
a total isospin zero, $\tau =0$, and all the possible states expressed in the
basis $\left| n_{1},n_{0},n_{-1}\right) $ are equivalent within a normalization
factor. The raising (lowering)
generators $\tau _{\pm 1}$ acting $\left( 2\tau +1\right) $ times on the
lowest $\left| n,\tau ,-\tau \right\rangle $ (highest $\left| n,\tau ,\tau
\right\rangle $) weight state give all the basis states of the respective $
\tau $-representation according to the result
\begin{equation}
\tau _{\pm 1}\left| n,\tau ,i\right\rangle =\sqrt{\frac{\left( \tau \mp
i\right) \left( \tau \pm i+1\right) }{2\Omega _{j}}}\left| n,\tau ,i\pm %
1\right\rangle .
\end{equation}

\item  The reduction chain ${Sp(4)\supset {U}^{0}{(2)\supset }SU}^{0}{(2)}%
\otimes {U}_{\tau _{0}}{(1)}$ introduces another labelling scheme for the
basis states, namely, $\left|i, {2(}n_{1}+n_{-1})_{\max },n\right).$ The
quantum numbers that specify the states are the eigenvalue $i$ of $\ \tau
_{0},$ $i=$ $-\Omega _{j},-\Omega _{j}+1,...,\Omega _{j},$ the seniority
quantum number $\ \nu ={2(}n_{1}+n_{-1})_{\max }=2\left| i\right| $, and the
eigenvalue of the operator $A_{0}^{0}=-(\Omega _{j}-\left| i\right| ),-(%
\Omega _{j}-\left| i\right| )+1,...,(\Omega _{j}-\left| i\right| )$ (\ref{gu0})$.$ The first invariant of ${U}^{0}
{(2)}$ decomposes the spaces ${\cal E}_{j}^{+}$ into a direct sum of
eigensubspaces of the operator $\ \tau _{0}$ at each of its fixed values
(\ref{iev}). These subspaces are represented by the columns of {\bf Table 1.}
The operator $A_{0}^{0}$ (\ref{gu0}) does not differ essentially from the
first invariant operator $N$ of $U_{N}(1)$ and it further reduces the columns
of {\bf Table 1} to the cells. The seniority quantum number differs between
two states of one and the same $i$ and $n,$ but different coupling scheme,
and it is introduced by the eigenvalues of the second Casimir operator for
this subgroup:
\begin{equation}
\begin{array}{l}
C_{2}(SU^{0}(2))\left| n_{1},n_{0},n_{-1}\right) = \\
\\
=\frac{2\Omega _{j}{-2(}n_{1}+n_{-1})_{\max }}{2}(\frac{2\Omega _{j}{-2(}%
n_{1}+n_{-1})_{\max }}{2}{+1)}\left| n_{1},n_{0},n_{-1}\right) .
\end{array}
\end{equation}
This is a scheme for coupling particles of the two different kinds $\{\sigma
=1,$ $\sigma ^{\prime }=-1\}$ with $n_{1}=0,$ or $n_{-1}=0,$ or both $
n_{1}=n_{-1}=0.$ These states are the last ones in each of the cells in the
{\bf Table 1}. The additional quantum number, $\nu ={2(}n_{1}+n_{-1})_{\max }
$, is the maximum number of the remaining pairs coupled as $\{\sigma =1,$ $
\sigma ^{\prime }=1\}$ or $\{\sigma =-1,$ $\sigma ^{\prime }=-1\}.\ $ In that
limit, the Casimir operator can be expressed in terms of the eigenvalue of
the first order invariant of ${U}^{0}{(2),}$
\begin{equation}
C_{2}(SU^{0}(2))\left| n_{1},n_{0},n_{-1}\right) =(\Omega _{j}-\left|
i\right| )(\Omega _{j}-\left| i\right| +1)\left| n_{1},n_{0},n_{-1}\right) .
\end{equation}
The raising and lowering generators of the subgroup ${SU}^{0}{(2)}$ act
along the columns in the following way:
\begin{eqnarray}
{A}_{+1}^{0}\left| n_{1},n_{0},n_{-1}\right)  &=&\left|
n_{1},n_{0}+1,n_{-1}\right) ,  \label{Azero_ca} \\
{A}_{-1}^{0}\left| n_{1},n_{0},n_{-1}\right)  &=&n_{0}\left( 1-\frac{%
2(n_{-1}+n_{1})+n_{0}-1}{2\Omega _{j}}\right) \left|
n_{1},n_{0}-1,n_{-1}\right) .  \nonumber
\end{eqnarray}

In each column $i$, ${A}_{+1}^{0}$ starts from the lowest$\ ${weight
}state
$\left| n_{1},n_{0},0\right) $ or $\left| 0,n_{0},n_{-1}\right) ,$ $
n_{0}=0,n_{\pm 1}=0,1,...,\left| i\right| $ and gives all the basis states
within a $\tau _{0}$-representation with $n_{0}=1,2,...,2(\Omega _{j}-\left|
i\right| )$. Similarly, ${A}_{-1}^{0}$ gives the basis states of the
representation of the subgroup under consideration, starting with the highest
weight state $n_{0}=2(\Omega _{j}-\left| i\right| ),$ for each $i.$

The normalized basis states,
\begin{equation}
\left| n_{1},n_{0},n_{-1}\right\rangle =\frac{1}{{\cal N}_{0}\left(
n_{1},n_{0},n_{-1}\right) }\left| n_{1},n_{0},n_{-1}\right) ,
\end{equation}
can be derived from (\ref{Azero_ca}). For the three types of states in this
reduction, the normalization coefficients are given by
\begin{equation}
\begin{array}{l}
{\cal P}_{0}^{2}\left( n_{1},n_{0},n_{-1}\right)
=n_{0}!\prod_{k=0}^{n_{0}-1}\left( 1-\frac{2(n_{-1}+n_{1})+k}{2\Omega _{j}}
\right) , \\
{\cal N}_{0}\left( 0,n_{0},0\right) ={\cal P}_{0}\left( 0,n_{0},0\right) ,
\\
{\cal N}_{0}\left( n_{1},n_{0},0\right) ={\cal P}_{0}\left(
n_{1},n_{0},0\right) {\cal N}\left( n_{1}\right) , \\
{\cal N}_{0}\left( 0,n_{0},n_{-1}\right) ={\cal P}_{0}\left(
0,n_{0},n_{-1}\right) {\cal N}\left( n_{-1}\right) ,
\end{array}
\label{norm0}
\end{equation}
where the lowest weight state $(n_{0}=0$ and $n_{\mp 1}=0)$ in each
representation can be normalized recursively,
\begin{equation}
{\cal N}^{2}\left( n_{\pm 1}\right) =n_{\pm 1}!\prod_{l=0}^{n_{\pm
1}-1}\left( 1-\frac{l}{\Omega _{j}}\right) .  \label{norm}
\end{equation}

\item  The other reduction is described again by the invariants of the
subgroups in the reduction chain: ${Sp(4)\supset U}^{\pm }{(2)\ \supset \ SU}
^{\pm }{(2)\otimes U(1)}_{N_{\mp }}.$ Here the labeling is $\left| n_{\mp
1}, {n}_{0}={n}_{0\max },n_{\pm 1}\right)$. First the spaces ${\cal
E}_{j}^{+}$ are decomposed by means of the first order invariants $N_{\mp }$
of the respective subalgebras to the subspaces defined by the conditions
$(2n_{\mp 1}+n_{0})$ $=0,1,...,2\Omega _{j}$ and represented by the left
(right) diagonals in {\bf Table 1}. The action of the Casimir operator on the
states
\begin{equation}
\begin{array}{l}
C_{2}(SU^{\pm }(2))\left| n_{1},n_{0},n_{-1}\right)=  \\
\\
=\frac{\Omega _{j}{-n}
_{0\max }}{2}(\frac{\Omega _{j}{-n}_{0\max }}{2}{+1)}\left|
n_{1},n_{0},n_{-1}\right),
\end{array}
\end{equation}
provides the $su(2)$ quantum number $d=({\Omega -n}_{0\max
})/2.$ The seniority quantum number ${n}_{0\max }$ is the maximum number of
remaining pairs that can be formed by coupling particles of different types,
here
${n}_{0\max }=\{0$ or $ 1\}.\ $The basis states are of the form $\left(
{{{A_{1}}}^{\dagger }}
\right) ^{n_{1}}\left( {{{A_{-1}}}^{\dagger }}\right) ^{n_{-1}}\left|
0\right\rangle $ and $\left( {{{A_{1}}}^{\dagger }}\right) ^{n_{1}}{{{A_{0}}}
^{\dagger }}\left( {{{A_{-1}}}^{\dagger }}\right) ^{n_{-1}}\left|
0\right\rangle $ and they are placed first in each cell in {\bf Table1.}
Furthermore, the operators $D_{0}^{\pm }$ (\ref{cgpm}), which are equivalent
within constants to the operators $N_{\pm }$, give the respective projection
of $d$: $D_{0}^{\pm } \left| n_{1},n_{0},n_{-1}\right) =\frac{1}{2}(2n_{\pm
}+n_{0\max }-\Omega  _{j}) \left| n_{1},n_{0},n_{-1}\right)=d_{0}^{\pm }
\left| n_{1},n_{0},n_{-1}\right)$. The diagonals are decomposed to the cells
belonging to them and defined by the conditions
$d_{0}^{\pm }=-d,-d+1,...,d.$

The raising and lowering generators of ${SU}^{\pm }{(2)}$ act along the
left/right diagonals:
\begin{eqnarray}
{A}_{+1}^{\pm }\left| n_{1},n_{0},n_{-1}\right)  &=&\left| n_{\pm
1}+1,n_{0},n_{\mp 1}\right) ,  \label{A+-_ca} \\
{B}_{-1}^{\pm }\left| n_{1},n_{0},n_{-1}\right)  &=&n_{\pm 1}\left( 1-\frac{
n_{\pm 1}+n_{0}-1}{\Omega _{j}}\right) \left| n_{\pm _{1}}-1,n_{0},n_{\mp
1}\right) .  \nonumber
\end{eqnarray}
Starting from the respective lowest or highest weight states, they generate
all the states belonging to $IUR$s of the ${U}^{\pm }{(2)\ }$subgroups of $%
Sp(4).$

The normalized basis states,
\begin{equation}
\left| n_{1},n_{0},n_{-1}\right\rangle =\frac{1}{{\cal N}_{\pm }\left(
n_{1},n_{0},n_{-1}\right) }\left| n_{1},n_{0},n_{-1}\right) ,
\end{equation}
can be derived from (\ref{A+-_ca}). For the two types of states ${n}_{0}=\{0$
or $1\}$ in this reduction, the normalization coefficients are given by
\begin{equation}
{\cal N}_{\pm }^{2}\left( n_{1},n_{0},n_{-1}\right)
=n_{1}!n_{-1}!\prod_{l=0}^{n_{1}-1}\left( 1-\frac{n_{0}+l}{\Omega _{j}}
\right) \prod_{l=0}^{n_{-1}-1}\left( 1-\frac{n_{0}+l}{\Omega _{j}}\right)
\label{norm+-}
\end{equation}
where the results (\ref{norm+-}) are consistent with (\ref{norm}) for $
n_{0}=0$ and with the lowest weight state in each representation ($n_{\pm
1}=0$) normalized recursively:
\begin{equation}
{\cal N}^{2}\left( n_{\mp 1},n_{0}\right) =n_{0}!\prod_{l=0}^{n_{0}-1}\left(
1-\frac{2n_{\mp 1}+l}{2\Omega _{j}}\right) n_{\mp 1}!\prod_{l=0}^{n_{\mp
1}-1}\left( 1-\frac{l}{\Omega _{j}}\right) .
\end{equation}
\end{enumerate}

\section{q-Deformations of the fermion realization of sp(4)}

Consider $q$-deformed creation and annihilation operators $\alpha _{m,\sigma
}^{\dagger }$ and $\alpha _{m,\sigma },\ m=-j,-j+1,...,j,\ \sigma =\pm 1,$
for a particle of type $\sigma $ in a state of a total angular momentum $j,$
with projection $m$ on the $z$ axis. The Hermitian conjugation relation is
defined as $(\alpha _{m,\sigma }^{\dagger })^{*}=\alpha _{m,\sigma }$. 

\subsection{q-Deformed transformation of the fermion operators}

There is a general class of functions, which transform the classical
operators into deformed ones \cite{fiore1,kuda}. We use the transformation
\begin{equation}
\alpha _{m,\sigma }=\theta ^{\frac{N_{\sigma }}{2}}c_{m,\sigma }\ ,\qquad
\alpha _{m,\sigma }^{\dagger }=c_{m,\sigma }^{\dagger }\bar{\theta}^{\frac{
N_{\sigma }}{2}} , \label{f->qf}
\end{equation}
where $\theta $ is a complex number with amplitude $\left| \theta \right|
=q, $ $q$ a real number, and $N_{\sigma }=\sum_{m}N_{m,\sigma }$ are
the classical number operators. The transformation of (\ref{cfc})
leads to the anticommutation relations for the $q$-deformed fermion
operators,
\begin{equation}
\alpha _{m,\sigma }\alpha _{m,\sigma }^{\dagger }+q\alpha _{m,\sigma
}^{\dagger }\alpha _{m,\sigma }=q^{N_{\sigma }},  \label{carq}
\end{equation}
and the identities

\begin{eqnarray}
\sum_{m}\alpha _{m,\sigma }^{\dagger }\alpha _{m,\sigma } &=&N_{\sigma
}q^{N_{\sigma }-1}, \\
\sum_{m}\alpha _{m,\sigma }\alpha _{m,\sigma }^{\dagger } &=&\left( 2\Omega
_{j}-N_{\sigma }\right) q^{N_{\sigma }}.  \nonumber
\end{eqnarray}

The raising and lowering generators of the respective deformed $Sp(4)$ group
are given as in the classical case (\ref{prlg}-\ref{cg}) but in terms of the
$q$-deformed fermion operators:

\begin{equation}
{{{F_{\sigma ,\sigma ^{\prime }}}}}=\xi _{\sigma ,\sigma ^{\prime
}}\sum_{m=-j}^{j}{{(-1)}^{j-m}}\alpha _{m,\sigma }^{\dagger }\alpha
_{-m,\sigma ^{\prime }}^{\dagger }{=}F{{{_{\sigma ^{\prime },\sigma
}=(G_{\sigma ,\sigma ^{\prime }})}}}^{\dagger },  \label{qg1}
\end{equation}
\begin{equation}
{{{G_{\sigma ,\sigma ^{\prime }}}}}=\xi _{\sigma ,\sigma ^{\prime
}}\sum_{m=-j}^{j}{{(-1)}^{j-m}}\alpha {_{-m,\sigma }}\alpha {_{m,\sigma
^{\prime }}},  \label{qg2}
\end{equation}
and

\begin{equation}
E{_{_{1,-1}}}=\eta \sum_{m=-j}^{j}\alpha {{{_{m,1}^{\dagger }}}}\alpha {\
_{m,-1}},\quad E{_{_{-1,1}}}=\eta \sum_{m=-j}^{j}\alpha {{{_{m,-1}^{\dagger }%
}}}\alpha {_{m,1},}  \label{dnpg}
\end{equation}
where the constants are defined in (\ref{shell scalling constants}). The
operator $F_{\sigma \sigma }$ $(G_{\sigma \sigma })$ creates (destroys) a
$q$-deformed pair of particles of the same kind.

The remaining two Cartan generators $N_{\sigma },\sigma =\pm 1,$ used in the
deformed commutation relations (\ref{carq}), are not deformed. The
transformation (\ref{f->qf}) yields the following relations between the
deformed (\ref{qg1}-\ref{dnpg}) and the classical operators (\ref{prlg}-\ref
{cg}):
\begin{equation}
{{{F_{\sigma ,\sigma }=A}}}_{\sigma ,\sigma }{\bar{\theta}^{N_{\sigma }+
\frac{1}{2}},\quad {{F_{1,-1}=A}}}_{1,-1}{\bar{\theta}^{\frac{N}{2}}},
\end{equation}
\begin{equation}
{{{G_{\sigma ,\sigma }=}}\theta ^{N_{\sigma }+\frac{1}{2}}{B}_{\sigma
,\sigma },\quad {G{_{1,-1}=}}\theta ^{\frac{N}{2}}{B}}_{1,-1},
\end{equation}
and
\begin{equation}
E{_{_{1,-1}}}=D_{1,-1}{\theta ^{\frac{N_{-}-1}{2}}\bar{\theta}^{\frac{N_{+}}{%
2}}},\quad E{_{_{-1,1}}}=D_{-1,1}{\theta ^{\frac{N_{+}-1}{2}}\bar{\theta}^{%
\frac{N_{-}}{2}}.}
\end{equation}
Since there is a smooth transformation that depends on the Cartan generators
of $sp(4)$ only and maps the $q$-deformed operators
\begin{equation}
F_{k}^{\dagger }=F_{\frac{1}{2}}^{\dagger }{{{_{(\sigma +\sigma ^{\prime })}}%
}}\equiv {F}_{\sigma ,\sigma ^{\prime }}\quad {(}G_{-k}=G_{-\frac{1}{2}}{{{\
_{(\sigma +\sigma ^{\prime })}}}}\equiv {G}_{\sigma ,\sigma ^{\prime }}),
\label{qg}
\end{equation}
\[
\sigma ,\sigma ^{\prime }=\pm 1,k=\frac{1}{2}(\sigma +\sigma ^{\prime
})=0,\pm 1,
\]
to the classical vectors $A{{{_{0,\pm 1}^{\dagger }(B{{{\ _{0,\mp 1}}}}),}}}$
the $q$-deformed states are equivalent within a phase to the classical ones (%
\ref{csA})${.}$ All the relations revert back to the classical formulae in
the limit $q\rightarrow 1$. The important reduction of $sp_{q}(4)$ algebra
to compact $u_{q}(2)$ subalgebra can be used again to obtain classification
schemes for the basis states.

\begin{enumerate}
\item  The subalgebra $u_{q}(2)$ of $sp_{q}(4)$ is closed by the number
preserving Weyl generators (\ref{dnpg}) and $N_{\sigma },\sigma =\pm 1,$
defined as:
\begin{equation}
\begin{array}{ll}
T_{+}{\equiv E{_{_{1,-1}}},} & T_{0}\equiv {{\tau }_{0}}=\frac{N{_{1}}-N_{-1}%
}{2}, \\
T_{-}{\equiv E{_{_{-1,1}}},} & {N}={N_{1}}+{N_{-1}.}
\end{array}
\label{qg_tau}
\end{equation}

The generators $T_{0},T_{\pm 1}$ and ${N}$ satisfy the commutation
relations:
\begin{equation}
\begin{array}{ll}
\lbrack T_{1},T_{-1}]=\frac{T_{0}}{\Omega _{j}}q^{N-1}, & {[}T_{0},T_{\pm 1}{
]}=\pm T_{\pm 1}, \\
\left[ N,T_{\pm 1}\right] =0, & \left[ N,T_{0}\right] =0.
\end{array}
\label{ttrc}
\end{equation}
and the second invariant of $u_{q}(2)$ is
\begin{equation}
C_{2}=N(N+1)-{\bf T}^{2}.
\end{equation}
The $q$-deformed operator ${\bf T}^{2}$ is defined by
\begin{eqnarray}
{\bf T}^{2} &=&\frac{2\Omega _{j}}{2}%
(T_{1}T_{-1}+T_{-1}T_{1})+T_{0}T_{0}q^{N-1}  \nonumber \\
&=&2\Omega _{j}T_{-1}T_{1}+T_{0}\left( T_{0}+1\right) q^{N-1},
\end{eqnarray}
and is related to the classical Casimir operator of $SU^{\tau }(2)$ (\ref
{i2}) by
\begin{equation}
{\bf T}^{2}={\bf \tau }^{2}q^{N-1}.
\end{equation}

Thus, the eigenvalues of the Casimir operator are deformed by a phase factor
$q^{n-1}$ and the eigenvectors are the classical basis states, $\left|
n,\tau ,i\right\rangle$.

\item  The other subgroup $U_{q}^{0}(2)$ is generated by the operators:
\begin{equation}
K_{+1}^{0}\equiv F{{{_{1,-1},\quad }}}K_{-1}^{0}\equiv G{{{_{1,-1},\quad }}
}K_{0}^{0}\equiv \frac{N}{2}-\Omega _{j}  \label{qg_zero}
\end{equation}
and $T_{0}$ (\ref{qg_tau}), which is the first order invariant$.$ The
generators of ${SU}_{q}^{0}{(2)}$ commute in the following way:
\begin{equation}
\lbrack K_{+1}^{0}{,K_{-1}^{0}]}_{-2}=\frac{K_{0}^{0}}{\Omega _{j}}q^{N{-2}
},\quad \left[ K_{0}^{0},K_{\pm 1}^{0}\right] =\pm K_{\pm 1}^{0},
\end{equation}
where the $q$-commutator is defined as
\begin{equation}
\left[ A,B\right] _{k}=AB-q^{k}BA.
\end{equation}

The second order Casimir invariant of ${su}_{q}^{0}{(2)}$ is given by
\begin{eqnarray}
C_{2}(SU^{0}(2)) &=&\frac{2\Omega _{j}}{2}%
(q^{2}K_{+1}^{0}K_{-1}^{0}+K_{-1}^{0}K_{+1}^{0})q^{-N}+\left(
K_{0}^{0}\right) ^{2}  \nonumber \\
&=&2\Omega _{j}K_{-1}^{0}K_{+1}^{0}q^{-N}+K_{0}^{0}\left( K_{0}^{0}+1\right).
\end{eqnarray}

\item  The two mutually complementary subalgebras $su_{q}^{+}(2)$ and $
su_{q}^{-}(2)$ of the algebra $sp_{q}(4)$ are given by the $q$-deformed
operators
\begin{equation}
F_{+1}^{\pm }=F_{\pm 1,\pm 1},G_{-1}^{\pm }=G_{\pm 1,\pm 1}  \label{qg_+-}
\end{equation}
and the non-deformed Cartan operators
\begin{equation}
E_{0}^{\pm }=\frac{N_{\pm 1}}{2}-\frac{\Omega _{j}}{2}.  \label{qg_cl}
\end{equation}
According to the reduction chain ${Sp}_{q}{(4)\supset SU}_{q}^{\pm }{(2)}
\otimes {U(1)}_{N_{\mp }}$, $N_{\mp }$ commute with the operators\ $F_{+1}^{
\pm },G_{-1}^{\pm },$ $E_{0}^{\pm }$ , which close the ${su_{q}^{\pm
}(2)}$ algebra:
\begin{equation}
\lbrack F_{+1}^{\pm }{,G_{-1}^{\pm }]}_{-4}=\frac{2E_{0}^{\pm }}{
\Omega _{j}}{\ q^{2N_{\pm 1}{-3}},}
\end{equation}
\begin{equation}
\lbrack N_{\pm 1},F_{+1}^{\pm }{]}{=2}F_{+1}^{\pm }{,\quad }[N_{\pm 1}{\
,G_{-1}^{\pm }]=-2G_{-1}^{\pm }}.
\end{equation}
The corresponding Casimir invariant is
\begin{eqnarray}
C_{2}(SU^{\pm }(2)) &=&\frac{\Omega _{j}}{2}(q^{4}F_{+1}^{\pm }G_{-1}^{\pm %
}+G_{-1}^{\pm }F_{+1}^{\pm })q^{-2N_{\pm 1}{-1}}+\left( E_{0}^{\pm }\right)
^{2}  \nonumber \\
&=&\Omega _{j}G_{-1}^{\pm }F_{+1}^{\pm }q^{-2N_{\pm 1}{-1}}+E_{0}^{\pm %
}\left( E_{0}^{\pm }+1\right) .
\end{eqnarray}
\end{enumerate}

A similar $q$-deformation is based on the transformation $\alpha _{m,\sigma
}=\theta ^{-\frac{N_{\sigma }}{2}}c_{m,\sigma },$ which yields the same
relations and identities as above, but with the exchange $q\rightarrow
q^{-1}.$ When $\theta $ is real and positive the deformation parameter is $
\theta \equiv q$.

\subsection{q-Deformation of the anticommutation relations of the fermion
operators}

Consider another set of $q$-deformed Hermitian conjugate operators
$\alpha _{m,\sigma }^{\dagger }$ and $\alpha _{m,\sigma },\ (\alpha _{m,\sigma
}^{\dagger })^{*}=\alpha _{m,\sigma },\ m=-j,-j+1,...,j,\ \sigma =\pm 1.$
Let the $q$-deformed anticommutation relation holds for every $\sigma $ and
$m$ in the form \cite{hay,ritsch}:

\begin{equation}
\alpha _{m,\sigma }\alpha _{m,\sigma }^{\dagger }+q^{\pm 1}\alpha _{m,\sigma
}^{\dagger }\alpha _{m,\sigma }=q^{\pm N_{m,\sigma }},  \label{usual f->qf}
\end{equation}
where $N_{m,\sigma }=$ $c{{{_{m,\sigma }^{\dagger }}}}c{_{m,\sigma }}$ and $
N_{\sigma }=\sum_{m=-j}^{j}N_{m,\sigma }$ are the classical number operators
(\ref{Nc}). Their action on the deformed fermion operators is defined as in
the classical case (\ref{NcCom}):
\begin{equation}
\begin{array}{ll}
\lbrack N_{\sigma },\alpha _{m,\sigma ^{\prime }}^{\dagger }]=\delta
_{\sigma ,\sigma ^{\prime }}\alpha _{m,\sigma ^{\prime }}^{\dagger } &
[N_{\sigma },\alpha _{m,\sigma ^{\prime }}]=-\delta _{\sigma ,\sigma
^{\prime }}\alpha _{m,\sigma ^{\prime }}
\end{array}
,\ \sigma ,\sigma ^{\prime }=\pm 1.
\end{equation}
In the previous section we showed that if the transformation function
(\ref{f->qf}) is used, the anticommutation relations of the deformed fermion
operators (\ref{carq}) depend not only on a single term $N_{m,\sigma }$ as in
(\ref{usual f->qf}) but rather on the total sum $N_{\sigma }.$ The same
dependence, along with the requirement that the deformation is performed only
on the $\sigma $ index, defines:

\begin{equation}
\alpha _{m,\sigma }\alpha _{m,\sigma }^{\dagger }+q^{\pm 1}\alpha _{m,\sigma
}^{\dagger }\alpha _{m,\sigma }=q^{\pm \frac{N_{\sigma }}{2\Omega _{j}}}.
\label{midcr}
\end{equation}
Using both anticommutation relations, it follows that $\alpha _{m,\sigma
}^{\dagger }\alpha _{m,\sigma }=[\frac{N_{\sigma }}{2\Omega _{j}}],$ where $%
[X]=\frac{q^{X}-q^{-X}}{q-q^{-1}},$ which leads to
\begin{equation}
\sum_{m}\alpha _{m,\sigma }^{\dagger }\alpha _{m,\sigma }=2\Omega _{j}[\frac{
N_{\sigma }}{2\Omega _{j}}]  \label{qNc}
\end{equation}
and
\begin{equation}
\sum_{m}\alpha _{m,\sigma }\alpha _{m,\sigma }^{\dagger }=2\Omega _{j}[1-%
\frac{N_{\sigma }}{2\Omega _{j}}].  \label{qNc-}
\end{equation}
In the limit $q\rightarrow 1,$ presuming $\alpha _{m,\sigma }^{\pm
}\rightarrow $ $c_{m,\sigma }^{\pm }$ as well, (\ref{qNc}, \ref{qNc-})
revert back to the classical formulas for $N_{\sigma }$ (\ref{Nc}, \ref{Nc-}
). This justifies the introduction of the weight coefficient $\omega \equiv
1/(2\Omega _{j})$ in (\ref{midcr}). The remaining anticommutation relations
for the $q$-deformed operators can be chosen from among various possibilities
\cite{pusz, fiore2}:

\begin{equation}
\begin{array}{ll}
\{\alpha _{m,\sigma },\alpha _{m^{\prime },\sigma }^{\dagger }\}_{q^{\pm
1}}=q^{\pm \frac{N_{\sigma }}{2\Omega _{j}}}\delta _{m,m^{\prime }}, &
\{\alpha _{m,\sigma },\alpha _{m^{\prime },\sigma ^{\prime }}^{\dagger
}\}=0,\sigma \neq \sigma ^{\prime }, \\
\{\alpha _{m,\sigma }^{\dagger },\alpha _{m^{\prime },\sigma ^{\prime
}}^{\dagger }\}=0, & \{\alpha _{m,\sigma },\alpha _{m^{\prime },\sigma
^{\prime }}\}=0,
\end{array}
\label{qfcr}
\end{equation}
where the $q$-anticommutator is given by $\left\{ A,B\right\}
_{k}=AB+q^{k}BA.$

The set of generators for this realization of the deformed $sp_{q}(4)$
algebra is defined as in (\ref{qg1}-\ref{dnpg}), but in terms of the $q$%
-deformed creation and annihilation operators $\alpha _{m,\sigma }^{\dagger
} $ $(\alpha _{m,\sigma }),$ fulfilling anticommutation relations (\ref
{qfcr}). The Cartan generators $N_{\pm 1}$ remain the classical number
operators. These ten operators generate the $q$-deformed $Sp_{q}(4)$ group
and its subgroup structure is investigated in analogy with the classical
case.

\begin{enumerate}
\item  The subgroup $U_{q}(2)$ of $Sp_{q}(4)$ is generated by the number
preserving Weyl operators (\ref{dnpg}) and $N_{\sigma },\sigma =\pm 1,$ as
well as by the equivalent set of the operators $T{_{0,\pm 1}}$ and $N$ (\ref
{qg_tau}). These operators satisfy the commutation relations:
\begin{equation}
\begin{array}{ll}
{\lbrack }T_{+},T_{-}{]}=[2\frac{{{T}_{0}}}{2\Omega _{j}}], & {[}T_{0},T_{%
\pm }{]}=\pm T_{\pm }, \\
\left[ N,T_{\pm }\right] =0, & \left[ N,T_{0}\right] =0.
\end{array}
\label{crq}
\end{equation}

The operators $T{_{0,\pm 1}}$ close an algebra $su_{q}(2)\sim so_{q}(3)$.
The number operator $N$ plays the role of the first order invariant of $
U_{q}(2)=SU_{q}(2)\otimes U(1)$. The second order Casimir operator of the
subgroup
$SU_{q}(2)$ is:
\begin{eqnarray}
{\bf T}^{2} &=&\frac{2\Omega _{j}}{2}(T_{+}T_{-}+T_{-}T_{+}+\left[ \omega
T_{0}\right] \left[ T_{0}+1\right] _{\omega }+\left[ \omega T_{0}\right]
\left[ T_{0}-1\right] _{\omega })  \nonumber \\
&=&2\Omega _{j}(T_{-}T_{+}+\left[ \omega T_{0}\right] \left[ T_{0}+1\right]
_{\omega }).  \label{qC2tau}
\end{eqnarray}

Here $\left[ X\right] _{\omega }=\frac{q^{\omega X}-q^{-\omega X}}{q^{\omega
}-q^{-\omega }}$ and the following identity has been used:
\begin{equation}
\left[ \omega T_{0}\right] \left[ T_{0}+1\right] _{\omega }-\left[ \omega
T_{0}\right] \left[ T_{0}-1\right] _{\omega }=\left[ 2\omega T_{0}\right] .
\label{qid}
\end{equation}
The Casimir operator coincides with the classical one in the limit $%
q\rightarrow 1$ (\ref{i2}).

\item  The other  $u_{q}^{0}(2)$ subalgebra is:
\begin{equation}
\begin{array}{ll}
\lbrack K_{+1}^{0},K_{-1}^{0}]=[2\frac{K_{0}^{0}}{2\Omega _{j}}], & \left[
K_{0}^{0},K_{\pm 1}^{0}\right] =\pm K_{\pm 1}^{0}, \\
\lbrack T_{0},K_{\pm 1}^{0}]=0, & [T_{0},K_{0}^{0}]=0,
\end{array}
\label{cu2q0}
\end{equation}

where the generators are defined in (\ref{qg_zero}).

The operator $T_{0}$ (\ref{qg_tau}) commutes with the generators of $%
su_{q}^{0}(2)$ (\ref{cu2q0}) and acts as a first order invariant of $%
u_{q}^{0}(2)=su_{q}^{0}(2)\oplus u_{T_{0}}^{0}(1).$ The operators $\left\{
K_{k}^{0}\right\} ,k=0,\pm 1$\ couple $q$-deformed particles of two
different kinds. The second order Casimir operator of the subgroup
$SU_{q}^{0}(2)$ is given by
\begin{eqnarray}
{\bf C}_{2}(SU_{q}^{0}(2)) &=&\frac{2\Omega _{j}}{2}%
(K_{+1}^{0}K_{-1}^{0}+K_{-1}^{0}K_{+1}^{0}+  \nonumber \\
&&\left[ \omega K_{0}^{0}\right] \left[ K_{0}^{0}+1\right] _{\omega }+\left[
\omega K_{0}^{0}\right] \left[ K_{0}^{0}-1\right] _{\omega }) \\
&=&2\Omega _{j}(K_{-1}^{0}K_{+1}^{0}+\left[ \omega K_{0}^{0}\right] \left[
K_{0}^{0}+1\right] _{\omega }),  \nonumber
\end{eqnarray}
which coincides with the classical invariant $($\ref{CSU0}$)$ in the limit $
q\rightarrow 1.$

\item  The two mutually complementary subalgebras $su_{q}^{+}(2)$ and $
su_{q}^{-}(2)$ of the algebra $sp_{q}(4)$ are given by the $q$-deformed
operators (\ref{qg_+-}) and the non-deformed Cartan operators (\ref{qg_cl}).
They have the following commutation relations:
\begin{eqnarray}
&&
\begin{tabular}{l}
$\left[ F_{+1}^{\pm },G_{-1}^{\pm }\right] =\rho _{\pm }[4\omega E_{0}^{\pm
}],$%
\end{tabular}
\nonumber \\
&&
\begin{tabular}{ll}
$\left[ E_{0}^{\pm },F_{+1}^{\pm }\right] =F_{+1}^{\pm },$ & $\left[ E_{0}^{
\pm },G_{-1}^{\pm }\right] =-G_{-1}^{\pm },$
\end{tabular}
\label{cu2q+-}
\end{eqnarray}
with $\rho _{\pm }=\frac{q^{\pm 1}+q^{\pm \frac{1}{2\Omega _{j}}}}{2}.$ It
is again true that each of the generators $\left\{
F_{+1}^{+},G_{-1}^{+},E_{0}^{+}\right\} $ of $SU_{q}^{+}(2)$ commutes with
all the generators of the other $SU_{q}^{-}(2)$ subgroup $\left\{
F_{+1}^{-},G_{-1}^{-},E_{0}^{-}\right\} $. The first order invariants $
N_{\mp 1}$ of $u_{q}^{\pm }(2)$ give the extension of $su_{q}^{\pm }(2)$ to
the subgroup $u_{q}^{\pm }(2)=$ $su_{q}^{\pm }(2)\oplus u^{\mp }(1).$ The
operator $F_{+1}^{\pm }$ $(G_{-1}^{\pm })$ creates (destroys) a $q$-deformed
pair of particles of the same kind. The Casimir invariant of the subgroup $
SU_{q}^{\pm }(2)$ is:
\begin{eqnarray}
{\bf C}_{2}(SU_{q}^{\pm }(2)) &=&\frac{\Omega _{j}}{2}(F_{+1}^{\pm }G_{-1}^{
\pm }+G_{-1}^{\pm }F_{+1}^{\pm }+\rho _{\pm }\left[ 2\omega E_{0}^{\pm
}\right] \left[ E_{0}^{\pm }+1\right] _{2\omega }+  \nonumber \\
&&\rho _{\pm }\left[ 2\omega E_{0}^{\pm }\right] \left[ E_{0}^{\pm %
}-1\right] _{2\omega }) \\
&=&\Omega _{j}(G_{-1}^{\pm }F_{+1}^{\pm }+\rho _{\pm }\left[ 2\omega E_{0}^{%
\pm }\right] \left[ E_{0}^{\pm }+1\right] _{2\omega }).  \nonumber
\end{eqnarray}
The useful identity (\ref{qid}) now has the form
\begin{equation}
\left[ 2\omega E_{0}^{\pm }\right] \left[ E_{0}^{\pm }+1\right] _{2\omega
}-\left[ 2\omega E_{0}^{\pm }\right] \left[ E_{0}^{\pm }-1\right] _{2\omega
}=\left[ 4\omega E_{0}^{\pm }\right].
\end{equation}

The Casimir operator coincides with the classical one (\ref{CSU+-}) in the
limit $q\rightarrow 1$.
\end{enumerate}

The $q$-deformed symplectic algebra reverts back to the classical limit for
the rest of the commutation relations between its generators (\ref{qg}):
\begin{eqnarray}
\lbrack F_{l}^{\dagger },G_{k}]_{2(k-l)} &=&\frac{\varphi _{l,k}}{2\sqrt{
\Omega _{j}}}T_{l+k}q^{(l-k)\omega N_{l-k}},\quad l+k\neq 0,  \nonumber \\
\lbrack T_{l},F_{k}^{\dagger }]_{k-l} &=&\frac{\chi _{l,k}}{\sqrt{\Omega _{j}%
}}F_{l+k}^{\dagger }q^{-l\omega N_{-l}},\quad l\neq 0,  \label{qcr1} \\
\lbrack T_{l},G_{k}]_{k-l} &=&-\frac{\phi _{l,k}}{\sqrt{\Omega _{j}}}%
G_{l+k}q^{l\omega N_{l}},\quad l\neq 0,  \nonumber \\
\lbrack T_{0},F_{k}^{\dagger }] &=&kF_{k}^{\dagger },\quad
[T_{0},G_{k}]=kG_{k},  \nonumber
\end{eqnarray}
where the constants are defined as follows:
\begin{eqnarray}
\varphi _{\pm 1,0} &=&2q^{\mp 2}\rho _{\pm } ,\ \varphi _{0,\pm 1}=2q^{\pm
(2+\frac{1}{2\Omega _{j}})}\rho _{\mp } ,\ \varphi _{\pm 1,\pm 1}=0,
\nonumber \\
\chi _{1,-1} &=&\rho _{-},\  \chi _{-1,1}=\rho _{+},\ \chi
_{\pm 1,0}=1,\ \chi _{\pm 1,\pm 1}=0, \\
\ \phi _{1,-1} &=&q^{-1}\rho _{-},\ \phi _{-1,1}=q\rho _{+} ,\ \phi _{\pm
1,0}=q^{\mp 1},\ \phi _{\pm 1,\pm 1}=0.  \nonumber
\end{eqnarray}
Another set of the same commutation relations can be obtained, which is
symmetric with respect to the exchange $q\leftrightarrow q^{-1}$ $:$
\begin{eqnarray}
\lbrack F_{l}^{\dagger },G_{k}] &=&\frac{1}{2\sqrt{\Omega _{j}}}\frac{1}{%
\left[ 2\right] }T_{l+k}\Psi _{lk}(N_{l-k}),\quad l+k\neq 0,  \nonumber \\
\lbrack T_{l},F_{k}^{\dagger }] &=&\frac{1}{2\sqrt{\Omega _{j}}}\frac{1}{%
\left[ 2\right] }F_{l+k}^{\dagger }\Psi _{l0}(N_{k}),\quad l,k\neq 0,
\nonumber \\
\lbrack T_{l},G_{k}] &=&-\frac{1}{2\sqrt{\Omega _{j}}}\frac{1}{\left[
2\right] }G_{l+k}\Psi _{0k}(N_{-k}),\quad l,k\neq 0,  \label{qcr} \\
\lbrack T_{l},F_{0}^{\dagger }]_{\frac{\left[ 2\right] }{2}} &=&\frac{1}{2%
\sqrt{\Omega _{j}}}F_{l}^{\dagger }\left( q^{\omega N_{-l}}+q^{-\omega
N_{-l}}\right) ,\quad l\neq 0,  \nonumber \\
\lbrack T_{l},G_{0}]_{\frac{\left[ 2\right] }{2}} &=&-\frac{1}{2\sqrt{\Omega
_{j}}}G_{l}\left( q^{\omega N_{l}}+q^{-\omega N_{l}}\right) ,\quad l\neq 0,
\nonumber
\end{eqnarray}
where the functions $\Psi _{lk}(N_{p})$ are defined in the following way:
\begin{equation}
\Psi _{lk}(N_{p})=\left\{
\begin{array}{c}
q^{\omega N_{p}}+q^{-\omega N_{p}}+q^{\omega (N_{p}+1)-1}+q^{-\omega
(N_{p}+1)+1},\quad k=0 \\
q^{\omega N_{p}-1}+q^{-\omega N_{p}+1}+q^{\omega (N_{p}-1)}+q^{-\omega
(N_{p}-1)},\quad l=0
\end{array}
\right. .
\end{equation}

The realization of $sp_{q}(4)$ introduced here is consistent with the
algebra of the Chevalley generators of ${\sl U}_{q}(SO(5)),$ which is given
in \cite{aach}. The comparison of the commutation relations yields for the
first triplet of the generators corresponding to the long root 1:
\begin{equation}
SU_{q}^{\tau }(2):(e_{1},f_{1},h_{1})\leftrightarrow (T_{+},T_{-},\omega
T_{0}),  \label{1}
\end{equation}
and for the short root 2:
\begin{equation}
SU_{q}^{-}(2):(e_{2},f_{2},h_{2})\leftrightarrow (\frac{F_{+1}^{-}}{\sqrt{[2]%
}},\frac{G_{-1}^{-}}{\sqrt{[2]}},\omega E_{0}^{-}).  \label{2}
\end{equation}
The renormalization of the generators of the second triplet is introduced so
that (\ref{cu2q+-}) can be written in the standard $su_{q}(2)$ form:
\begin{equation}
\left[ \frac{F_{+1}^{\pm }}{\sqrt{[2]}},\frac{G_{-1}^{\pm }}{\sqrt{[2]}}%
\right] =\rho _{\pm }\frac{[4\omega E_{0}^{\pm }]}{[2]}=\rho _{\pm }[2\omega
E_{0}^{\pm }]_{2}.
\end{equation}
The rest of the commutation relations of both triplets are consistent within
the parameter $\omega .$ Comparing (\ref{qcr1}) with the other four
generators, we obtain

\begin{equation}
(e_{3}^{+},f_{3}^{+})\leftrightarrow (F_{0}q^{-\omega
N_{-1}},G_{0}q^{-\omega N_{-1}})  \label{3}
\end{equation}
and
\begin{equation}
(e_{4}^{+},f_{4}^{+})\leftrightarrow (F_{+1}^{+}q^{-\omega
N_{-1}},G_{-1}^{+}q^{-\omega N_{-1}}),  \label{4}
\end{equation}
which are determined up to an overall multiplicative constant factor. The
results prove the isomorphism of the
$q$-fermion realization of $sp_{q}(4)$ and all possible representations
of its standard $SU_{q}(2)$ subgroup to the triplets of the Chevalley
generators associated with the shorter and longer roots of ${\sl U}
_{q}(SO(5)) $.

\subsection{Action space of the fermion realization of sp$_{q}$(4)}

In general, the $q$-deformed fermion operators (\ref{midcr}) act as in the
classical case in a finite metric space {\rm \ }${\cal E}_{j}$ for each
particular
$j$-level, with a vacuum $|0\rangle \ $defined by $\alpha _{m,\sigma }$ $
|0\rangle =0.$ The scalar product in ${\cal E}_{j}$ is chosen in such a way
that $\alpha _{m,\sigma }^{\dagger }$ is a Hermitian conjugate to $\alpha
_{m,\sigma }:(\alpha _{m,\sigma }^{\dagger })^{*}=\alpha _{m,\sigma },$ \ and
$\langle 0|0\rangle =1.$ In general the $q$-deformed states are different from
the classical ones, but reduce to the classical ones in the limit
$q\rightarrow 1.$

The $q$-deformed creation (annihilation) operators $F_{k}^{\dagger }{
,k=0,\pm 1}$ (\ref{qg}) are components of a tensor of rank $1$ with respect
to the subgroup $SU_{q}^{T}(2)$ (\ref{qcr}). These operators create a pair
of $q $-fermions coupled to a total angular momentum $J=0$ and a total
isospin $T=1$. Analogous to the classical limit, a set of vectors that
span each space ${\cal E}_{j}^{+}$ in the $q$-deformed case can be chosen to
be of the form

\begin{equation}
\left| n_{1},n_{0},n_{-1}\right) _{q}=\left( F{{{_{1}^{\dagger }}}}\right)
^{n_{1}}\left( F{{{_{0}^{\dagger }}}}\right) ^{n_{0}}\left( {{{
F_{-1}^{\dagger }}}}\right) ^{n_{-1}}\left| 0\right\rangle .  \label{csF}
\end{equation}
The basis is obtained by orthonormalization of (\ref{csF}). The index $q$
will be dropped from the notation for the basis states in the following cases
which treat only the deformed space.

As in the classical case, $\Omega _{j}$ labels the representation for each
particular $j$-shell. The basis states are uniquely specified by the
classification schemes which use the $su_{q}\left( 2\right) $ subalgebras and
the relevant Cartan generators. In the $q$-deformed case the Cartan generators
of $Sp_{q}\left( 4\right) $ can be chosen to be the nondeformed operators
$N_{\pm 1}$ or their equivalent set of operators $N$ and $T_{0}$ $\equiv \tau
_{0}$ (\ref{qg_tau}). The eigenvalues of these operators that label the basis
states coincide with the ones in the classical case and the example of {\bf
Table 1} can still be used. The quantum numbers provided by the eigenvalues of
the $q$-deformed Casimir invariants have to be taken in the limit
$q\rightarrow 1.$

We briefly list the reduction chains and compare them to their classical
counterparts in order to emphasize the similarity and differences between
them. The basis states together with the second order Casimir operators and
their eigenvalues are often used in the physical applications. It is in this
sense that their $q$-deformation may lead to some interesting new results.

\begin{enumerate}
\item  In the limit $q\rightarrow 1,$ the second order Casimir operator,
$ {\bf T}^{2}$, of the $SU_{q}^{T}(2)$ subgroup has the eigenvalues:
\begin{equation}
{\bf T}^{2}\left| n,T,i\right\rangle _{q}\mathrel{\mathop{\rightarrow
}\limits_{q\rightarrow 1}}T(T+1)\left| n,T,i\right\rangle ,
\end{equation}

where ${T=}\frac{\tilde{n}}{2},$ $\frac{\tilde{n}}{2}-2,...,1$ (odd) or $0$
(even), where $\tilde{n}=\min \left\{ n,4\Omega _{j}-n\right\} ,$ and $
i=-T,-T+1,...,T.$ In the deformed case the eigenvalues of ${\bf T}^{2}$
for the lowest and the highest weight states (\ref{qC2tau}) are
\begin{equation}
{\bf T}^{2}\left| n,T,\pm T\right\rangle _{q}=2\Omega _{j}\left[ \frac{1}{2
\Omega _{j}}\right] \left[ T\right] _{\omega }\left[ T+1\right] _{\omega
}\left| n,T,\pm T\right\rangle _{q}.
\end{equation}

\item  The reduction chain ${Sp}_{q}{(4)\supset SU}_{q}^{0}{(2)}\otimes {U}
_{q}{(1)}_{T_{0}}$ describes pairing between fermions of different types and
introduces the seniority quantum number ${2(}n_{1}+n_{-1})_{\max }$ in the
labelling scheme for the basis states, $\left| i,{2(}n_{1}+n_{-1})_{\max
},n\right).$ The eigenvalue of the second order Casimir operator for this $q
$-deformed subalgebra is given by
\begin{equation}
\begin{array}{l}
C_{2}(SU_{q}^{0}(2))\left| n_{1},n_{0},n_{-1}\right) = \\
=2\Omega _{j}\left[ \frac{1}{2\Omega _{j}}\right] \left[ \frac{2\Omega _{j}{
-2(}n_{1}+n_{-1})_{\max }}{2}\right] _{\omega }\left[ (\frac{2\Omega _{j}{-2(
}n_{1}+n_{-1})_{\max }}{2}{+1)}\right] _{\omega }\left|
n_{1},n_{0},n_{-1}\right) .
\end{array}
\end{equation}

Here again, the generators of the subalgebra ${su}_{q}^{0}{(2)}$ act along
the columns:
\begin{eqnarray}
{K}_{+1}^{0}\left| n_{1},n_{0},n_{-1}\right)  &=&\left|
n_{1},n_{0}+1,n_{-1}\right) ,  \nonumber \\
{K}_{-1}^{0}\left| n_{1},n_{0},n_{-1}\right)  &=&\left[ n_{0}\right] _{\frac{
1}{2\Omega _{j}}}\left[ 1-\frac{2(n_{-1}+n_{1})+n_{0}-1}{2\Omega _{j}}
\right] \left| n_{1},n_{0}-1,n_{-1}\right) ,  \nonumber \\
N\left| n_{1},n_{0},n_{-1}\right)  &=&2\left( n_{-1}+n_{1}+n_{0}\right)
\left| n_{1},n_{0},n_{-1}\right) .  \label{qKzero_ca}
\end{eqnarray}
The normalized basis states,
\begin{equation}
\left| n_{1},n_{0},n_{-1}\right\rangle =\frac{1}{{\cal M}_{0}\left(
n_{1},n_{0},n_{-1}\right) }\left| n_{1},n_{0},n_{-1}\right) ,
\end{equation}
can be derived from (\ref{qKzero_ca}). For the three types of states in this
reduction, the normalization coefficients are:
\begin{equation}
\begin{array}{l}
{\cal Q}_{0}^{2}\left( n_{1},n_{0},n_{-1}\right) =\left[ n_{0}\right]
_{\omega }!\prod_{k=0}^{n_{0}-1}\left[ 1-\frac{2(n_{-1}+n_{1})+k}{2\Omega
_{j}}\right] , \\
{\cal M}_{0}\left( 0,n_{0},0\right) ={\cal Q}_{0}\left( 0,n_{0},0\right) ,
\\
{\cal M}_{0}\left( n_{1},n_{0},0\right) ={\cal Q}_{0}\left(
n_{1},n_{0},0\right) {\cal M}\left( n_{1}\right) , \\
{\cal M}_{0}\left( 0,n_{0},n_{-1}\right) ={\cal Q}_{0}\left(
0,n_{0},n_{-1}\right) {\cal M}\left( n_{-1}\right) ,
\end{array}
\end{equation}
where the $q$-deformed factorial is defined by $\left[ A\right]
_{k}!=\left[ A\right] _{k}\left[ A-1\right] _{k}...1.$ The normalization
coefficients ${\cal M}\left( n_{\pm 1}\right) $ of the lowest weight state $
(n_{0}=0$ and $n_{\mp 1}=0)$ in each representation is derived by means of
the generators of the next reduction.

\item  The other reduction ${Sp}_{q}{(4)\supset U}_{q}{(2)}_{N_{\mp }}{\
\supset SU}_{q}^{\pm }{(2)\supset U}_{q}{(1)}_{N_{\pm }}$ introduces
deformation in the model of coupled fermions of the same kind. Here the
labeling is $\left| n_{\mp 1},{n}_{0}={n}_{0\max },n_{\pm 1}\right)$,
where ${n}_{0\max }=\{0$ or $1\}$ is the seniority quantum number. The
action of the Casimir operator on the states is given by
\begin{eqnarray}
C_{2}(SU_{q}^{\pm }(2))\left| n_{1},n_{0},n_{-1}\right)  &=&\rho _{\pm }{\ }
\Omega _{j}\left[ \frac{1}{\Omega _{j}}\right] \left[ \frac{\Omega _{j}{-n}
_{0\max }}{2}\right] _{2\omega }\times   \nonumber \\
&&\left[ \frac{\Omega _{j}{-n}_{0\max }}{2}{+1}\right] _{2\omega }\left|
n_{1},n_{0},n_{-1}\right) .
\end{eqnarray}

In the deformed case the action of the Casimir invariant of $SU_{q}^{+}(2)$
differs from that of the Casimir invariant of $SU_{q}^{-}(2)$ by the factor $
\rho _{+}/\rho _{-}.$ The generators of $su_{q}^{\pm }\left( 2\right) $
transform the states along the diagonals as
\begin{eqnarray}
{F}_{+1}^{\pm }\left| n_{1},n_{0},n_{-1}\right)  &=&\left| n_{\pm %
1}+1,n_{0},n_{\mp 1}\right) ,  \nonumber \\
{G}_{-1}^{\pm }\left| n_{1},n_{0},n_{-1}\right)  &=&\rho _{\pm }\left[ n_{
\pm 1}\right] _{\frac{1}{\Omega _{j}}}\left[ 1-\frac{n_{\pm 1}+n_{0}-1}{
\Omega _{j}}\right] \left| n_{\pm _{1}}-1,n_{0},n_{\mp 1}\right) ,  \nonumber
\\
N_{\pm }\left| n_{1},n_{0},n_{-1}\right)  &=&\left( 2n_{\pm 1}+n_{0}\right)
\left| n_{1},n_{0},n_{-1}\right) .  \label{qF+-_ca}
\end{eqnarray}
The normalized basis states,
\begin{equation}
\left| n_{1},n_{0},n_{-1}\right\rangle =\frac{1}{{\cal M}_{\pm }\left(
n_{1},n_{0},n_{-1}\right) }\left| n_{1},n_{0},n_{-1}\right) ,
\end{equation}
can be derived from (\ref{qF+-_ca}). For the two types of states ${n}_{0}=\{0
$ or $1\}$ in this reduction, the normalization coefficients are
\begin{eqnarray}
{\cal M}_{\pm }^{2}\left( n_{1},n_{0},n_{-1}\right)  &=&\rho _{+}\rho
_{-}\left[ n_{1}\right] _{2\omega }!\left[ n_{-1}\right] _{2\omega
}!\prod_{l=0}^{n_{1}-1}\left[ 1-\frac{n_{0}+l}{\Omega _{j}}\right] \times  
\nonumber \\ &&\prod_{l=0}^{n_{-1}-1}\left[ 1-\frac{n_{0}+l}{\Omega
_{j}}\right] ,
\end{eqnarray}
where for $n_{0}=0$ and $n_{\mp 1}=0$ it follows that
\begin{equation}
{\cal M}^{2}\left( n_{\pm 1}\right) =\rho _{\pm }\left[ n_{\pm 1}\right]
_{2\omega }!\prod_{l=0}^{n_{\pm 1}-1}\left( 1-\frac{l}{\Omega _{j}}\right) .
\end{equation}
\end{enumerate}

It is important to emphasize that this deformation may lead to basis states
whose content is very different from the classical case since there is no
known simple function that transforms the classical fermion operators
$c_{m,\sigma }^{\dagger }$ and $c_{m,\sigma }$ into the $q$-deformed ones
$\alpha _{m,\sigma }^{\dagger }$ and $\alpha _{m,\sigma }$. Smooth function
may not exist when the anticommutation relations (\ref{qfcr}) hold
simultaneously with both signs for one and the same $\sigma ,$ as they are
defined in (\ref{midcr}).

The deformed basis states are labeled by the classical eigenvalues of the
invariant operators of the reduction along each of the cases considered. The
matrix elements, particularly of the raising and lowering generators of
$sp_{q}(4)$ and the second order invariants, are also deformed which leads
to different results in physical applications. After obtaining the
correspondence between the $q$-fermion realization of $ sp_{q}(4)$ and the
Chevalley generators of ${\sl U}_{q}(SO(5))$ we can compare the two bases for
an irreducible representation $\Omega _{j}=\frac{1 }{4}n_{\max }$, which
corresponds to the representation $(n_{1},n_{2})$ at $
\ n_{1}=n_{2}=\frac{1}{4}n_{\max }$ \cite{aach}. In the classical and in the
$q$-deformed cases, the first basis considered in \cite{aach} is related to
the basis states (\ref{csA}) and (\ref{csF}). 

\section{Conclusion}

In this paper we consider a fermion realization of the $sp(4)$ algebra and
its deformations. The original algebra, as well as some of
its deformed realizations, act in the same finite space ${\cal E}_{j}.$ The
finiteness of the representations is due to the Pauli principle.

The deformed realization of $sp_{q}(4)$ is based on the standard
$q$-deformation of the two component Clifford algebra \cite{hay}, realized in
terms of creation and annihilation fermion operators. For the $sp(4)$ case
eight of the ten generators are deformed, the fermion number operators
$N_{1}$ and $N_{-1}$ and their linear combinations being the exceptions. The
deformed generators of $Sp_{q}(4)$ close different realizations of the compact
$u_{q}(2)$ subalgebra. The induced representations of each $u_{q}(2)$ are
reducible in the space ${\cal E}_{j}^{+}$ and decompose into irreducible
representations. In this way we obtain a full description of the irreducible
unitary representations of $U_{q}(2)$ of four different realizations of
$u_{q}(2)$: $u_{q}^{\tau }(2),\ u_{q}^{0}(2)$ and $u_{q}^{\pm }(2).$

Each reduction into compact subalgebras of $sp(4)$ and its deformations
affords the possibility of a description of a different physical model with
different dynamical symmetries. While within a particular deformation scheme
the basis states may either be deformed or not, the generators are always
deformed as is their action on basis states. With a view towards
applications, the additional parameter of the deformation gives a richer
variety of operators associated with observables, nondeformed as well as
deformed. In a Hamiltonian theory this implies a dependance of the matrix
elements on the deformation parameter, leading to the possibility of greater
flexibility and richer structures within the framework of algebraic
descriptions.


\begin{thebibliography}{99}
\bibitem{GoLiSp}  S. Goshen and H. J. Lipkin, {\it Spectroscopic and group
theoretical methods in physics}, Amsterdam (1968)

\bibitem{hecht}  K. T. Hecht, {\it Nucl. Phys.} {\bf A102} (1967) 11

\bibitem{bar}  A. O. Barut and R. Raczka, {\it Theory of Group
Representations and Applications, PWN, Warsaw }(1977)

\bibitem{Goli}  S. Goshen and H. J. Lipkin, {\it Ann. Phys.}(NY) {\bf 6}
(1959) 301

\bibitem{Roro}  G. Rosensteel and D. J. Rowe, {\it Phys. Rev. Lett.}{\bf \ 38
(} 1977) 10

\bibitem{pehe}  D. R. Peterson and K. T. Hecht, {\it Nucl. Phys.} {\bf A344}
(1980) 361

\bibitem{ker}  A. Kerman, {\it Ann. Phys.}(NY) {\bf 12} (1961) 300

\bibitem{cirevo}  O. Civitarese, M. Reboiro, P. Vogel, {\it Phys. Rev. C}
{\bf 56} (1997) 1840

\bibitem{klmap}  A. Klein, E. Marshalek, {\it Rev. Mod. Phys.} {\bf 63 }
(1991) 375.

\bibitem{nagedo}  P. Navratil, H. Geyer, J. Dobaczewski, {\it Nucl. Phys. }
{\bf A607} (1996) 23

\bibitem{Drinfeld} V. G. Drinfeld, {\it ``Quantum Groups"}, edited by A.
Gleason, in {\it Proceedings of the International Congress of Mathematicians},
Berkeley, 1986 (American Mathematical Society, Providence, 1987) Vol.1 798

\bibitem{liNoRu}  J. Lukierski, A. Nowicki, H. Ruegg, {\it Phys. Lett.}{\bf
\ B271 (} 1991) 321

\bibitem{aach}  B. Abdesselam, D. Arnaudon and A. Chakrabarti, {\it J. Phys.
A: Math. Gen. }{\bf 28 }(1995) 3701

\bibitem{hay}  T. Hayashi, {\it Commun. Math. Phys}. {\bf 127} (1990) 129

\bibitem{sha}  S. Shelly Sharma, N. Sharma, {\it Phys. Rev. C} {\bf 62}
(2000) 034314

\bibitem{debre}  J. P. Draayer, A. I. Georgieva and M. I. Ivanov, {\it J.
Phys. A: Math. Gen.} {\bf 34} (2001) 2999

\bibitem{spin-staistics theorem}  W. Pauli, {\it Phys. Rev.} {\bf 58} (1940)
716

\bibitem{flo}  B. H. Flowers, {\it Proc. Roy. Soc.} {\bf A212} (1952) 248

\bibitem{fiore1} G. Fiore, {\it J. Math. Phys.} {\bf 39}
(6) June 1998

\bibitem{kuda}  P. P. Kulish and E. V. Damaskinski, {\it J. Phys. A: Math.
Gen}. {\bf 23} (1990) L415

\bibitem{ritsch}  V. Rittenberg, M. Scheunert, {\it J. Math. Phys.} {\bf 33}
(2) February 1992

\bibitem{pusz}  W. Pusz, {\it Rep. Math. Phys}. {\bf 27} (1989) 349

\bibitem{fiore2} G. Fiore, {\it ``Clifford Algebras and their applications in
mathematical physics"} Vol.1: Algebra and Physics, ed. R. Ablamowicz, B.
Fauser, Birkhaeuser (2000) 269-282
\end{thebibliography}
\end{document}